\documentclass[prl,superscriptaddress,amsmath,amssymb,twocolumn]{revtex4-2}
\usepackage{natbib}
\usepackage{graphicx}         
\usepackage{dcolumn}          
\usepackage{bm}               
\usepackage{upgreek}
\usepackage{booktabs} 				
\usepackage{tabularx}
\usepackage{array}
\usepackage{ulem}
\usepackage{float}
\usepackage[colorlinks=true,urlcolor=blue,breaklinks=true]{hyperref}
\usepackage{xcolor}
\usepackage[a4paper,left=2cm,right=2cm]{geometry}

\begin{document}
\title{Fourier transformation based analysis routine for intermixed longitudinal and transversal hysteretic data for the example of a magnetic topological insulator}

\author{Erik Zimmermann}
\email{er.zimmermann@fz-juelich.de}
\affiliation{Peter Gr\"unberg Institut (PGI-9), Forschungszentrum J\"ulich, 52425 J\"ulich, Germany}
\affiliation{JARA-Fundamentals of Future Information Technology, J\"ulich-Aachen Research Alliance, Forschungszentrum J\"ulich and RWTH Aachen University, Germany}

\author{Michael Schleenvoigt}
\affiliation{Peter Gr\"unberg Institut (PGI-9), Forschungszentrum J\"ulich, 52425 J\"ulich, Germany}
\affiliation{JARA-Fundamentals of Future Information Technology, J\"ulich-Aachen Research Alliance, Forschungszentrum J\"ulich and RWTH Aachen University, Germany}

\author{Alina Rupp}
\affiliation{Peter Gr\"unberg Institut (PGI-9), Forschungszentrum J\"ulich, 52425 J\"ulich, Germany}
\affiliation{JARA-Fundamentals of Future Information Technology, J\"ulich-Aachen Research Alliance, Forschungszentrum J\"ulich and RWTH Aachen University, Germany}

\author{Gerrit Behner}
\affiliation{Peter Gr\"unberg Institut (PGI-9), Forschungszentrum J\"ulich, 52425 J\"ulich, Germany}
\affiliation{JARA-Fundamentals of Future Information Technology, J\"ulich-Aachen Research Alliance, Forschungszentrum J\"ulich and RWTH Aachen University, Germany}

\author{Jan Karthein}
\affiliation{Peter Gr\"unberg Institut (PGI-9), Forschungszentrum J\"ulich, 52425 J\"ulich, Germany}
\affiliation{JARA-Fundamentals of Future Information Technology, J\"ulich-Aachen Research Alliance, Forschungszentrum J\"ulich and RWTH Aachen University, Germany}

\author{Justus Teller}
\affiliation{Peter Gr\"unberg Institut (PGI-9), Forschungszentrum J\"ulich, 52425 J\"ulich, Germany}
\affiliation{JARA-Fundamentals of Future Information Technology, J\"ulich-Aachen Research Alliance, Forschungszentrum J\"ulich and RWTH Aachen University, Germany}

\author{Peter Sch\"uffelgen}
\affiliation{Peter Gr\"unberg Institut (PGI-9), Forschungszentrum J\"ulich, 52425 J\"ulich, Germany}
\affiliation{JARA-Fundamentals of Future Information Technology, J\"ulich-Aachen Research Alliance, Forschungszentrum J\"ulich and RWTH Aachen University, Germany}

\author{Hans L\"uth}
\affiliation{Peter Gr\"unberg Institut (PGI-9), Forschungszentrum J\"ulich, 52425 J\"ulich, Germany}
\affiliation{JARA-Fundamentals of Future Information Technology, J\"ulich-Aachen Research Alliance, Forschungszentrum J\"ulich and RWTH Aachen University, Germany}

\author{Detlev Gr\"utzmacher}
\affiliation{Peter Gr\"unberg Institut (PGI-9), Forschungszentrum J\"ulich, 52425 J\"ulich, Germany}
\affiliation{JARA-Fundamentals of Future Information Technology, J\"ulich-Aachen Research Alliance, Forschungszentrum J\"ulich and RWTH Aachen University, Germany}

\author{Thomas Sch\"apers}
\email{th.schaepers@fz-juelich.de}
\affiliation{Peter Gr\"unberg Institut (PGI-9), Forschungszentrum J\"ulich, 52425 J\"ulich, Germany}
\affiliation{JARA-Fundamentals of Future Information Technology, J\"ulich-Aachen Research Alliance, Forschungszentrum J\"ulich and RWTH Aachen University, Germany}

\hyphenation{}
\date{\today}

\begin{abstract}
We present a symmetrization routine that optimizes and eases the analysis of data featuring the anomalous Hall effect. This technique can be transferred to any hysteresis with (point-)symmetric behaviour. The implementation of the method is demonstrated exemplarily using intermixed longitudinal and transversal data obtained from a chromium-doped ternary topological insulator revealing a clear hysteresis. Furthermore, by introducing a mathematical description of the anomalous Hall hysteresis based on the error function precise values of the height and coercive field are determined.
\end{abstract}
\maketitle

\section{Introduction}
Magnetic topological insulators (MTIs) are characterized by their unique properties in band structure~\cite{tokura2019magnetic,chang2013experimental,chang2023colloquium}. MTIs can be formed by incorporating magnetic atoms such as chromium, vanadium or manganese into the topological insulator (TI) lattice~\cite{he2015quantum,ou2018enhancing,mogi2015magnetic,chang2015high,gong2019experimental,chang2013experimental}. Another possibility is proximity inducing magnetic moments with a ferromagnetic insulator in vicinity of a TI~\cite{zhu2018proximity}. These material platforms are considered to be suitable candidates for Majorana physics~\cite{adagideli2020time,hassler2020half,liu2018majorana,chen2018quasi,zeng2018quantum,beenakker2019deterministic}. Using MTIs the quantum anomalous Hall effect (QAHE) was first detected experimentally in 2013 by Chang et al.~\cite{chang2013experimental}. It is characterized by vanishing bulk conductance and a single spin-polarized edge mode that surrounds the sample~\cite{yu2010quantized,tokura2019magnetic}. When increasing the temperature or detuning the Fermi energy the (not quantized) anomalous Hall effect is observed in MTIs~\cite{onoda2003quantized,deng2020quantum,niu2019quantum,zhang2019experimental}.

\begin{figure}[hbtp]
\centering
\includegraphics[width=0.47\textwidth]{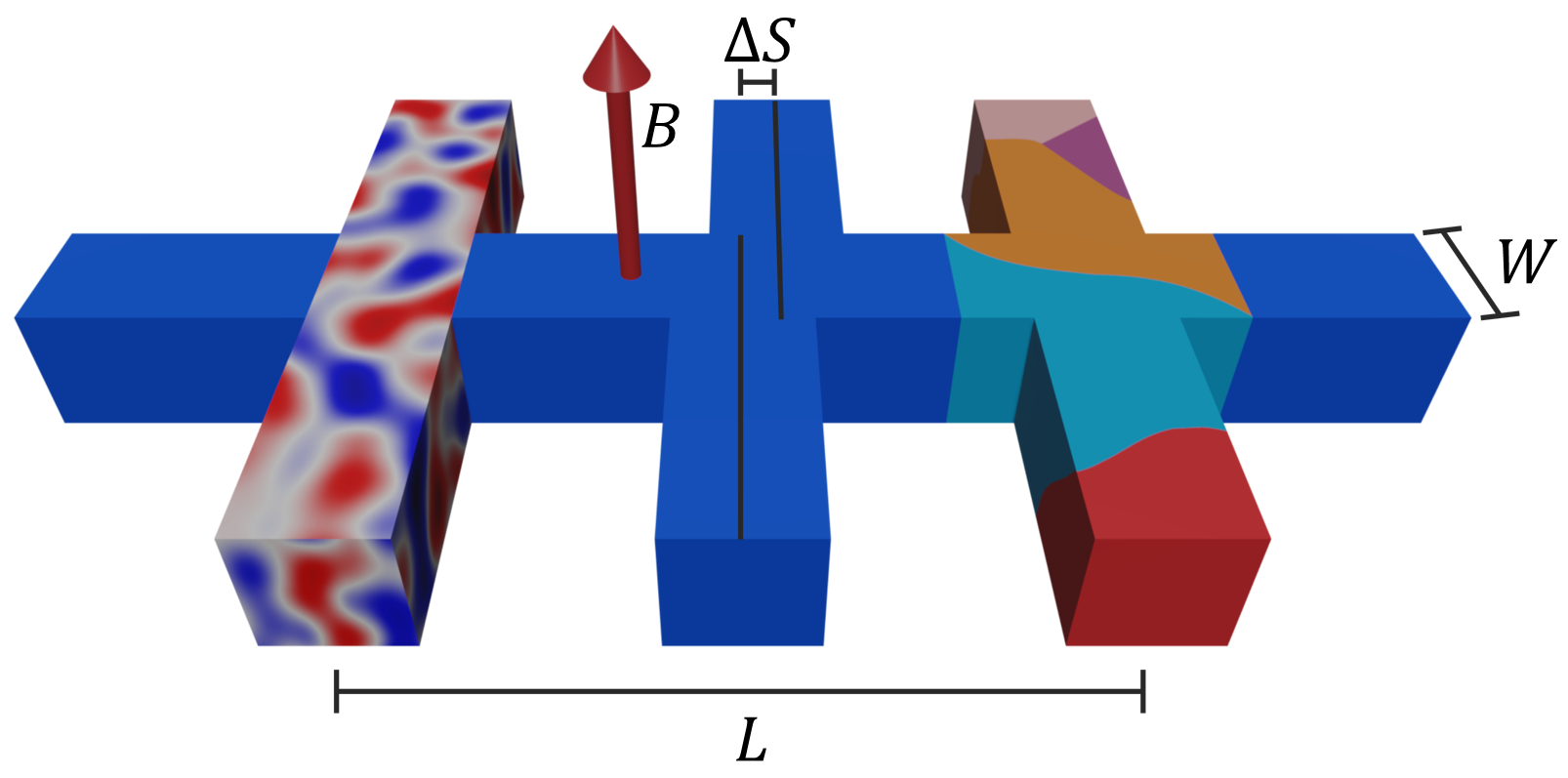}
\caption{Imperfect Hall bar as origin of intermixed longitudinal and transversal magnetotransport data. The orientation of the external magnetic field $B$ is indicated with an arrow. The examples of charge puddles (left), misaligned contacts (middle) or potential drops at grain boundaries (right) are sketched at opposing contact pairs.}\label{Fig_Versch_Kontakte}
\end{figure}

A common problem when dealing with experimentally gained magnetotransport data from Hall bars is the intermixing of longitudinal ($R_\text{xx}$) and transversal ($R_\text{xy}$) resistance data~\cite{laha2020magnetotransport,cao2021growth,watson2015dichotomy,budhani2021planar,shi2020anomalous}. Even when excluding errors, there are internal origins for an overlay of the signals that cannot be solved experimentally. Possible reasons are sketched in Fig.~\ref{Fig_Versch_Kontakte} on transversal contact pairs of a Hall bar. On the left contact pair inhomogeneous potential fluctuations due to charge puddles are shown~\cite{ito2022cancellation,knispel2017charge,borgwardt2016self}. In the middle a geometrical displacement ($\Delta S$) of the contacts is sketched that gains importance when approaching smaller structures dependent on the fabrication technique~\cite{chang2013experimental}. In this case, a longitudinal signal in the order of $\Delta S/L\cdot R_\text{xx}$ would contribute to the Hall signal $R_\text{xy}$. On the right contact pair possible grain boundaries causing inhomogeneous potential drops are depicted. The consequence of the presented mechanisms is comparable to the one of a diagonal measurement over the Hall bar~\cite{zimmermann2023universal}. Despite these irregularities, the data could hold valuable information about the sample.

In the following, the data processing of intermixed conventional and anomalous Hall data with respect to their symmetries is explained. Subsequently, the analysis routine is demonstrated using the data of the MTI Hall bar. Here, the focus lies on the symmetrization of the Hall data, the longitudinal data can be treated analogously.

\section{Data processing}
\subsection{Symmetries of Magnetic Topological Insulators}
When analyzing measurement data that shows an intermixing of longitudinal and Hall signal, symmetrizing the data (see supplementary material) is a common method to separate the signals from each other, as the longitudinal data is expected to show axial symmetry with respect to the ordinate while the transversal data is point symmetric. Thus, a fast Fourier transformation (FFT) can help separating the respective contributions. In order to do so, the signal is Fourier transformed to the frequency space giving complex values in general. For the longitudinal data all odd contributions given by the imaginary part are filtered out while for the transversal data all even ones given by the real part are omitted. Then, the remaining quantities are transformed back giving the unperturbed signal using inverse FFT.

\begin{figure}[H]
\centering
\includegraphics[width=0.47\textwidth]{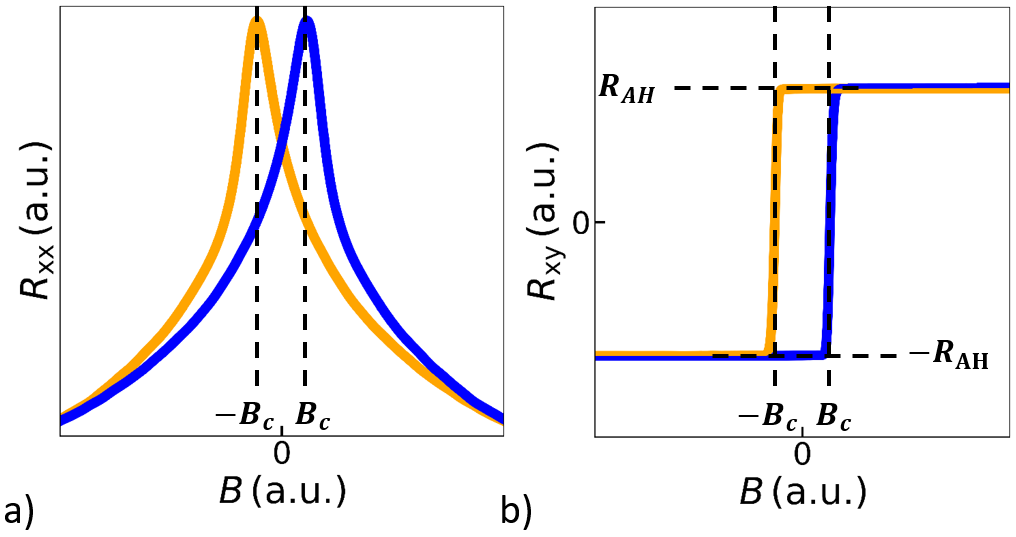}
\caption{Anomalous Hall effect. a) The longitudinal data $R_\text{xx}$ of two magnetic field sweeps is shown in arbitrary units (a.u.), where the blue data corresponds to a sweep from negative to positive magnetic field and the orange curve vice versa. The peak indicating zero total magnetic field is shifted in $B$ by the coercive magnetic field with respect to the zero position. b) The corresponding Hall data $R_\text{xy}$ shows a hysteresis with height $R_{\text{AH}}$ and width $B_c$.}\label{Fig_Beispiel_AHE_Hysterese}
\end{figure}

For MTIs showing the anomalous Hall effect or even the quantum anomalous Hall effect there are major differences. Figure~\ref{Fig_Beispiel_AHE_Hysterese} shows such signatures in the longitudinal (a) and Hall resistance (b) for two magnetic field sweeps: The blue curve shows data of the sweep from negative magnetic field to positive and the orange curve shows the data for the sweep in opposite direction \footnote{The data shown here is actual data from an MTI Hall bar.}. One can see that, compared to e.g. non-magnetic TIs, the sweeps differ around zero magnetic field. The origin are the magnetic moments $M$ introduced by the doping material that align in a ferromagnetic order with the external magnetic field $B$, so an internal magnetization inside the MTI is created~\cite{ye1999berry}, similar to diluted magnetic semiconductors~\cite{furdyna1988diluted}. At the coercive magnetic field $B_c$, the external magnetic field is strong enough to switch the orientation of the opposingly aligned internal magnetic moments. In the longitudinal signal around the switching point a shifted peak for each sweep direction is visible. In the Hall signal a hysteretic behaviour is seen that is identified by the coercive magnetic field $B_c$ and the height in resistance $R_{\text{AH}}$.

Due to the hysteretic behaviour, it becomes apparent, that the symmetrization procedure mentioned before is not directly transferable. Intuitively, one could think of shifting the data by $\pm B_c$ and do the same procedure, but especially when considering the longitudinal data one can see that this would only fit well for the peak position but not for larger magnetic fields. Indeed, one would also loose information about the individual curvature of the anomalous Hall hysteresis. A better but more complex solution is the reallocation of the data points in the data set to again find axial and point symmetry. 
\begin{figure}[hbtp]
\centering
\includegraphics[width=0.47\textwidth]{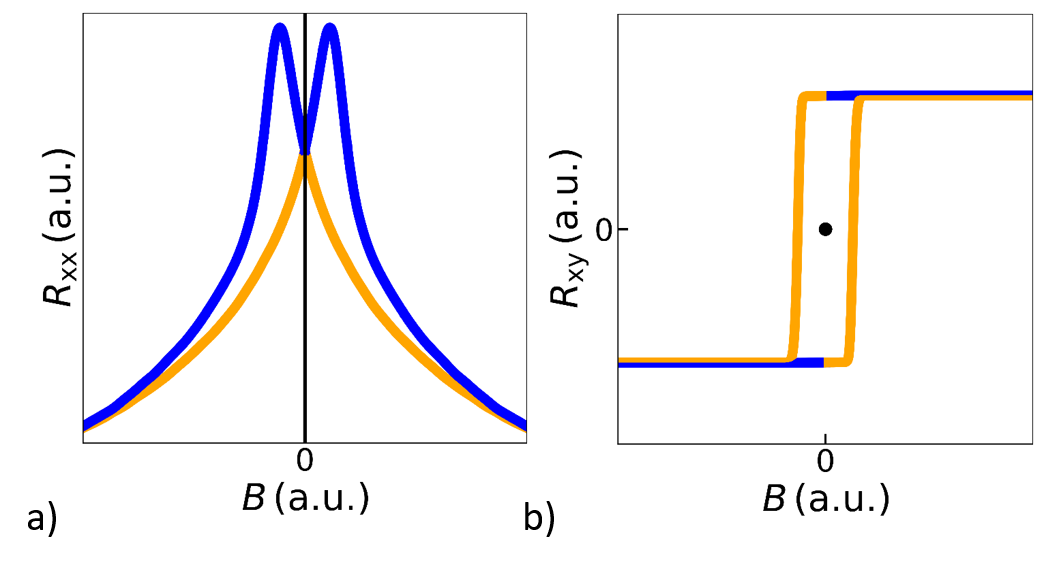}
\caption{Symmetries of MTIs. Compared to Fig.~\ref{Fig_Beispiel_AHE_Hysterese}, here, the colors mark the parts of the data sets that are supposed to be symmetric to each other. a) The longitudinal data of two magnetic field sweeps is shown in arbitrary units (a.u.). A symmetry axis at $B=0\,$T is shown. b) The corresponding Hall hysteresis shows a point symmetric behaviour. The symmetry point is indicated by a black dot at coordinate origin in the middle of the hysteresis.}\label{Fig_Beispiel_AHE_Hysterese_Neuaufteilung}
\end{figure}

Figure~\ref{Fig_Beispiel_AHE_Hysterese_Neuaufteilung} illustrates, how the data needs to be restructured in order to have data sets that show a certain symmetry. In Fig.~\ref{Fig_Beispiel_AHE_Hysterese_Neuaufteilung}~a) the longitudinal data is shown. Colored in blue and orange, respective data set pairs are marked that are symmetric with respect to the ordinate shown with a black mirror line. The Hall hysteresis is depicted in Fig.~\ref{Fig_Beispiel_AHE_Hysterese_Neuaufteilung}~b). Compared to the longitudinal signal that shows an axial symmetry with respect to the ordinate, the Hall data is point symmetric to the middle of the hysteresis at the coordinate origin as indicated by a black dot. When combining the orange and blue symmetries, it becomes apparent that the two sweep directions are not symmetric in itself but with respect to each other. Thus, for many systems merging both sweep directions to one data set is an equivalent alternative with the same symmetries. Having this new symmetry in mind, the Fourier symmetrization discussed above can be done.

\subsection{Mathematical Description of the Hall Hysteresis}
The parameters $B_c$ and $R_{\text{AH}}$ (cf. Fig.~\ref{Fig_Beispiel_AHE_Hysterese}) are taken as a suitable measure for the hysteresis. Thus, their accurate determination is crucial. Hence, after the symmetrization process described before, the cleared data set is resorted in the manner of the process depicted in Fig.~\ref{Fig_Beispiel_AHE_Hysterese}~b) and a fit is performed. Even after symmetrization the ideal data still consists of the hysteresis arising from the anomalous Hall effect superimposed by the classical Hall slope. As the slope may be non-negligible the data is corrected by the point symmetric conventional Hall slope determined far away from the hysteresis. Now, just the anomalous hysteresis is remaining.

As one can see in the transversal signal, the switching of the magnetic moments does not result in a step-like behaviour. Instead, the tails are slightly curved. The reason is that not all the magnetic moments are bound exactly the same way but are assumed to follow approximately a Gaussian distribution, for instance due to defects or due to the inhomogeneity of the energy at the edges of the sample. Therefore, e.g. a Heaviside step function as basic model would neglect these factors. Instead, we employed an error function as a mathematical model that describes the normalized integration of a Gaussian function. The complete model that describes the development of each branch of the hysteresis results in
\begin{equation}
    R(B)=R_{\text{AH}}\cdot \text{erf}(A(B\pm B_c))~.\label{eq_Erf_Hysteresis}
\end{equation}
Here, $R_{\text{AH}}$ scales the height of the normalized error function, the term $B\pm B_c$ takes the shift of the curve by the coercive magnetic field $B_c$ with respect to zero magnetic field into account and the parameter $A$ is a measure of the switching curvature. Further discussion regarding the model can be found in the supplementary material. Figure~\ref{Fig_Beispiel_AHE_Hysterese_Erf} shows the data from Fig.~\ref{Fig_Beispiel_AHE_Hysterese}~b). One least square fit for each side using equation~\ref{eq_Erf_Hysteresis} is performed and shown with a dotted line. One can see that the fits describe the data quite accurately. Moreover, three error functions with varying parameter $A$ are plotted in the insert to illustrate how the curvature of the function develops. In the following the method is demonstrated exemplarily using the intermixed anomalous Hall data of a chromium-based MTI.
\begin{figure}[hbtp]
\centering
\includegraphics[width=0.47\textwidth]{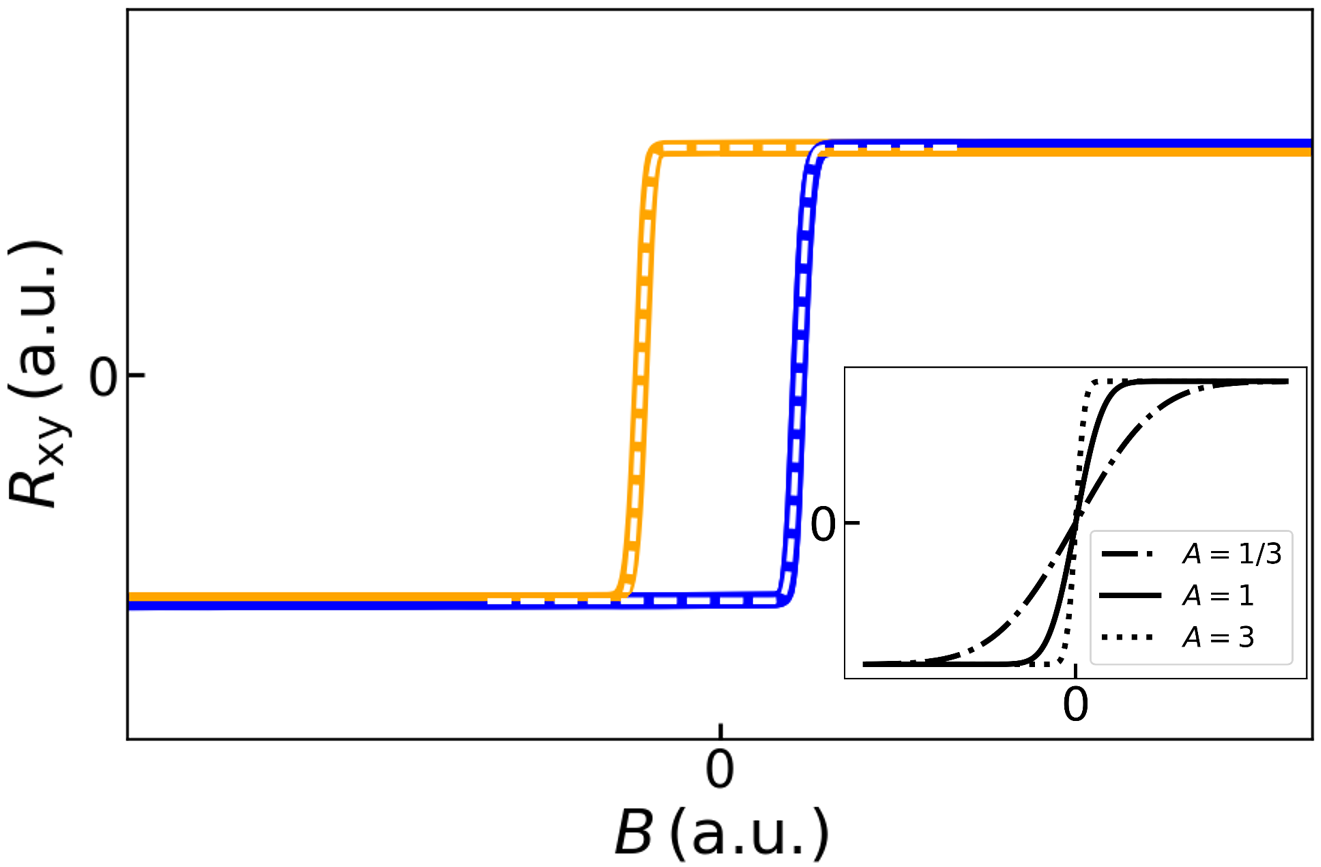}
\caption{Model for the Hall hysteresis. Using the least square method, the error function is fitted to the data taken from Fig.~\ref{Fig_Beispiel_AHE_Hysterese}~b). The results are shown with dotted lines. Error functions with different $A$ parameters are sketched in the insert.}\label{Fig_Beispiel_AHE_Hysterese_Erf}
\end{figure}
\begin{figure*}[hbtp]
\centering
\includegraphics[width=\textwidth]{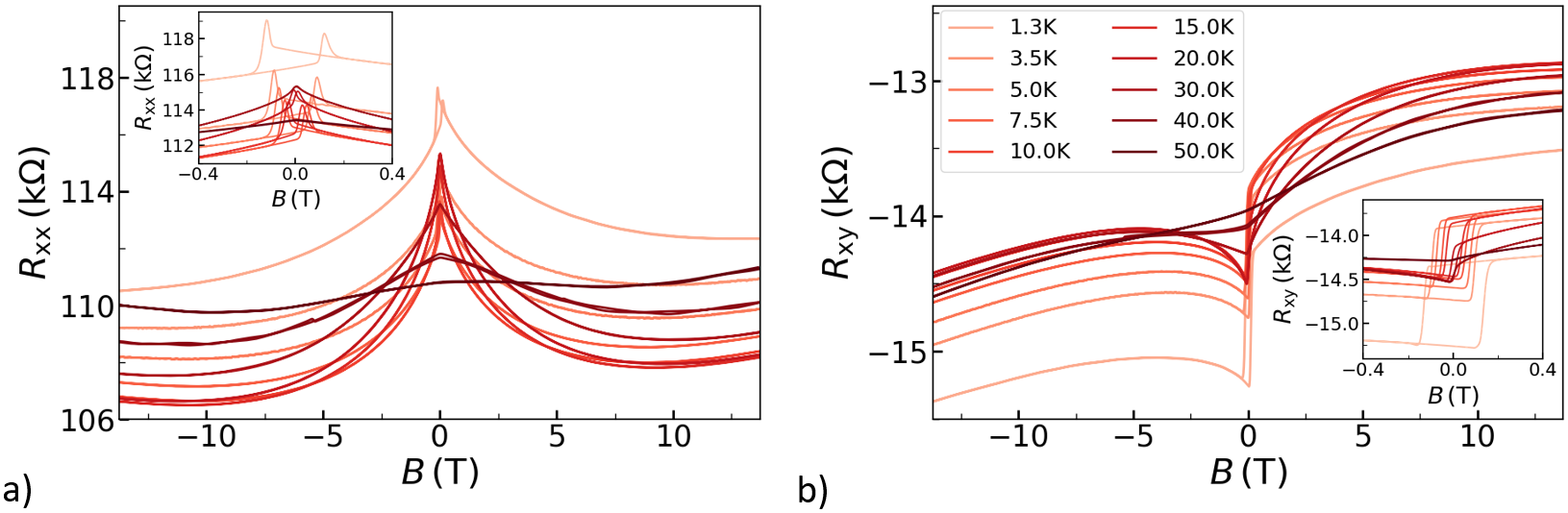}
\caption{Perturbed magnetoresistance signal of an MTI. The legend is shown in b). a) The raw longitudinal data recorded for different temperatures shows a small influence from the Hall signal. b) The corresponding raw Hall data is shown. The inserts show zoom-ins of precise measurements around zero magnetic field.}\label{Fig_Magn_Korrektur1}
\end{figure*}

\section{Exemplary analysis}
The following data is obtained from a $6\,\mu$m wide and $300\,\mu$m long MTI Hall bar with a stoichiometry of Cr$_{0.15}$Bi$_{0.35}$Sb$_{1.5}$Te$_{3}$. Further information on the fabrication and measurement technique is provided in the supplementary material. The sample produced slightly asymmetric data. Using the exact same setup, similar samples have been found to have no intermixing of the signal. Thus, the origin for the intermixing seen in this sample is attributed to an internal issue. The analysis is divided into two parts. First, the classical Hall effect is analyzed where high magnetic fields are beneficial for a destingued determination of the slope. After that, a precise measurement around the hysteresis feature is used for the investigation of the AHE.

\subsection{Classical Hall Analysis}
Figure~\ref{Fig_Magn_Korrektur1}~a) and b) show the longitudinal data and the raw Hall data, respectively. The peaks in the $R_\text{xx}$ signal are slightly affected by the Hall hysteresis. Furthermore, an offset in resistance between the values for high negative and high positive magnetic field is seen. In the transversal signal the intermixing of longitudinal data is even more pronounced due to the ratio of their magnitudes. In the $R_\text{xy}$ signal a dip around zero magnetic field, the typical (inverse) longitudinal curvature and an offset to negative values on the ordinate are observed as disruptive factors. To conclude, the signal is highly intermixed and especially the shape of the Hall signal does not correspond to the expectations.

The aim of this part of the analysis is to get an accurate estimation of the classical Hall slope. Therefore, the Hall data around zero magnetic field is excluded. This allows to handle the symmetrization of the data similar to the one of conventional TIs without need of restructuring the data sets for symmetry reasons. After removing the even contributions in the frequency space the resulting transversal signal is free from longitudinal contributions. With this data, a charge carrier concentration of $n_{\text{2D}}=1.86\cdot 10^{13}\,$cm$^{-2}$ and a mobility of $\mu=143$\,cm$^2$/Vs are derived from the slope at base temperature using classical Hall analysis and Drude theory. As a check, the value of the charge carrier concentration is also calculated to $n_{\text{2D}}=1.85\cdot 10^{13}\,$cm$^{-2}$ from the non-linear raw data. The values do not really differ, as the exclusion of the symmetric data arising from an intermixture of the longitudinal data does not effect the point symmetric, linear slope.
\begin{figure*}[hbtp]
\centering
\includegraphics[width=\textwidth]{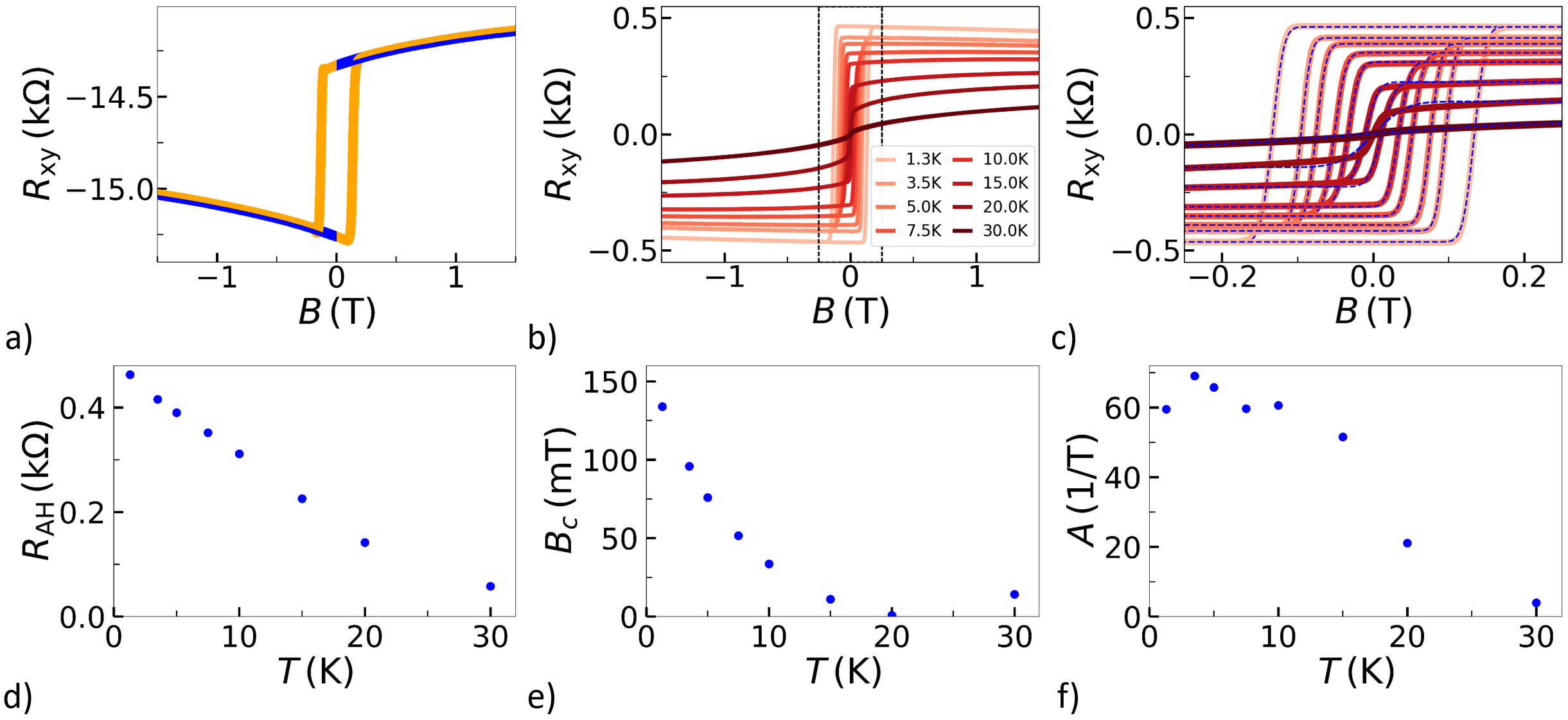}
\caption{Symmetrization and analysis of the AHE. The legend is shown in b). a) The redistribution for the symmetrization process of the signal at a base temperature of $1.3\,$K is shown. In b) the symmetrized data is shown. c) Fits of equation~\ref{eq_Erf_Hysteresis} are performed for all temperatures and the temperature dependences of the averaged fit parameters $R_\text{AH}$, $B_c$ and $A$ are plotted in d) - f), respectively.}\label{Fig_Magn_Korrektur2}
\end{figure*}

\subsection{Anomalous Hall Analysis}
The inserts of Fig.~\ref{Fig_Magn_Korrektur1}~a) and b) show precise measurements around zero magnetic field of the longitudinal and transversal data. Only the data up to $30\,$K is taken into account, as for higher temperatures no hysteretic behavior is observed.
First, the slope determined in the previous large magnetic field measurement is subtracted, as this is caused by the classical Hall effect and the purpose of the precise measurement around zero magnetic field is the determination of the anomalous Hall effect properties. As the slope is a point symmetric feature, it has to be excluded separately. In order to remove the contributions from $R_\text{xx}$ in the Hall signal, the data of both sweep directions is split at $B=0\,$T and restructured following the technique shown in Fig.~\ref{Fig_Beispiel_AHE_Hysterese_Neuaufteilung}. The redistribution of the values is indicated in Fig.~\ref{Fig_Magn_Korrektur2}~a) for base temperature. Next, the data is interpolated to ensure an equidistant spacing of the data points for the Fourier transformation. Making use of the point symmetry of the restructured $R_\text{xy}$ data, a Fourier analysis is performed that removes all axial symmetric contributions.

The result of the symmetrization process is shown in Fig.~\ref{Fig_Magn_Korrektur2}~b) for multiple temperatures. One can see that the signal is cleared from intermixed perturbations. The symmetric data that is excluded during the symmetrization process is plotted in Fig.~S2 in the supplementary material. A fitting of the point symmetric data using equation~\ref{eq_Erf_Hysteresis} is made for both branches of each hysteresis. The fits are plotted in Fig.~\ref{Fig_Magn_Korrektur2}~c) together with the data marked with a dashed box in Fig.~\ref{Fig_Magn_Korrektur2}~b). The fit describes the data well, but it can also be seen that a remaining curvature at small magnetic fields of the data measured at higher temperatures causes a slight difference between fit and data that results in an uncertainty in the value of $A$ for elevated temperatures. The corresponding fit parameters are shown in Fig.~\ref{Fig_Magn_Korrektur2}~d) - f). For the determination of the values the average of both fits for each temperature is taken. As the curves are similar, also due to the symmetrization, only differences below $0.01\,\%$ between the parameters of both fits are observed. For base temperature values of $R_\text{AH}=463\,\Omega$, $B_c=134\,$mT and $A=59.5\,$T$^{-1}$ are obtained. One can see that all parameters decrease with increasing temperatures. The decrease of $R_\text{AH}$ and $B_c$ indicate a decrease of the AHE towards the Curie temperature $T_c$. For $T=20\,$K the widths of the hysteresis is already close to zero so that for the measurement at $T=30\,$K no real value for $B_c$ could be determined. As the parameter $A$ scales inversely with the width of the transition, the transition region is broadened with increasing temperature. For $T\geq 20\,$K a larger decrease is observed as the bending of the curve also influences the parameter.

\section{Conclusion} 
In this article the analysis of intermixed conventional and anomalous Hall data was discussed. First, different reasons for intermixed data were listed. Then, a possible symmetrization process of conventional data using a FFT is shown. Thereafter, the existing axial symmetry of the longitudinal and the point symmetry of the transversal anomalous Hall data were discussed. A method for the restructuring of the data by splitting and recombining it at zero magnetic field is suggested in order to maintain the symmetries. This was followed by the symmetrization process using FFT. Furthermore, a mathematical description based on the error function is introduced in order to describe and fit the hysteresis.

Next, the data of an MTI sample was analyzed that showed intermixed anomalous Hall data. The method is carried out exemplarily for the transversal data of the MTI Hall bar. The result is a clear, symmetric anomalous Hall hysteresis where the height and width are precisely determined using the model based on the error function. Slight deviations in curvature from the fitting model are found for elevated temperatures as in addition to the approximately Gaussian distributed binding of the magnetic moments another temperature dependent component that is expected to be Fermi-Dirac distributed contributes.

Besides clearing the hysteretic data from perturbation, the presented approach offers a precise determination of the width and the height of the hysteresis that are comparable to the ones estimated from the raw data. This approach cannot only be used for magnetic field dependent MTI measurements but may be transferred easily to any perturbed hysteretic behaviour that is based on symmetries.

Finally, it is pointed out that the method needs to be handled with care as it may mimic symmetries also for data where no symmetry is expected. Thus, a close comparison between resulting data, raw data and underlying concepts always needs to be made.

\section{Acknowledgments}
We thank Herbert Kertz for technical assistance and Jonas Buchhorn for fruitful discussion. All samples have been prepared at the Helmholtz Nano Facility~\cite{albrecht2017hnf}. This work is funded by the Deutsche Forschungsgemeinschaft (DFG, German Research Foundation) under Germany's Excellence Strategy – Cluster of Excellence Matter and Light for Quantum Computing (ML4Q) EXC 2004/1 – 390534769, by the German Federal Ministry of Education and Research (BMBF) via the Quantum Futur project ‘MajoranaChips’ (Grant No. 13N15264) within the funding program Photonic Research Germany and by the QuantERA grant MAGMA via the German Research Foundation under grant 491798118.
\noindent 

\begin{thebibliography}{33}%
\makeatletter
\providecommand \@ifxundefined [1]{%
 \@ifx{#1\undefined}
}%
\providecommand \@ifnum [1]{%
 \ifnum #1\expandafter \@firstoftwo
 \else \expandafter \@secondoftwo
 \fi
}%
\providecommand \@ifx [1]{%
 \ifx #1\expandafter \@firstoftwo
 \else \expandafter \@secondoftwo
 \fi
}%
\providecommand \natexlab [1]{#1}%
\providecommand \enquote  [1]{``#1''}%
\providecommand \bibnamefont  [1]{#1}%
\providecommand \bibfnamefont [1]{#1}%
\providecommand \citenamefont [1]{#1}%
\providecommand \href@noop [0]{\@secondoftwo}%
\providecommand \href [0]{\begingroup \@sanitize@url \@href}%
\providecommand \@href[1]{\@@startlink{#1}\@@href}%
\providecommand \@@href[1]{\endgroup#1\@@endlink}%
\providecommand \@sanitize@url [0]{\catcode `\\12\catcode `\$12\catcode
  `\&12\catcode `\#12\catcode `\^12\catcode `\_12\catcode `\%12\relax}%
\providecommand \@@startlink[1]{}%
\providecommand \@@endlink[0]{}%
\providecommand \url  [0]{\begingroup\@sanitize@url \@url }%
\providecommand \@url [1]{\endgroup\@href {#1}{\urlprefix }}%
\providecommand \urlprefix  [0]{URL }%
\providecommand \Eprint [0]{\href }%
\providecommand \doibase [0]{http://dx.doi.org/}%
\providecommand \selectlanguage [0]{\@gobble}%
\providecommand \bibinfo  [0]{\@secondoftwo}%
\providecommand \bibfield  [0]{\@secondoftwo}%
\providecommand \translation [1]{[#1]}%
\providecommand \BibitemOpen [0]{}%
\providecommand \bibitemStop [0]{}%
\providecommand \bibitemNoStop [0]{.\EOS\space}%
\providecommand \EOS [0]{\spacefactor3000\relax}%
\providecommand \BibitemShut  [1]{\csname bibitem#1\endcsname}%
\let\auto@bib@innerbib\@empty
\bibitem [{\citenamefont {Tokura}\ \emph {et~al.}(2019)\citenamefont {Tokura},
  \citenamefont {Yasuda},\ and\ \citenamefont
  {Tsukazaki}}]{tokura2019magnetic}%
  \BibitemOpen
  \bibfield  {author} {\bibinfo {author} {\bibfnamefont {Y.}~\bibnamefont
  {Tokura}}, \bibinfo {author} {\bibfnamefont {K.}~\bibnamefont {Yasuda}}, \
  and\ \bibinfo {author} {\bibfnamefont {A.}~\bibnamefont {Tsukazaki}},\ }\href
  {\doibase https://doi.org/10.1038/s42254-018-0011-5} {\bibfield  {journal}
  {\bibinfo  {journal} {Nature Reviews Physics}\ }\textbf {\bibinfo {volume}
  {1}},\ \bibinfo {pages} {126} (\bibinfo {year} {2019})}\BibitemShut {NoStop}%
\bibitem [{\citenamefont {Chang}\ \emph {et~al.}(2013)\citenamefont {Chang},
  \citenamefont {Zhang}, \citenamefont {Feng}, \citenamefont {Shen},
  \citenamefont {Zhang}, \citenamefont {Guo}, \citenamefont {Li}, \citenamefont
  {Ou}, \citenamefont {Wei}, \citenamefont {Wang} \emph
  {et~al.}}]{chang2013experimental}%
  \BibitemOpen
  \bibfield  {author} {\bibinfo {author} {\bibfnamefont {C.-Z.}\ \bibnamefont
  {Chang}}, \bibinfo {author} {\bibfnamefont {J.}~\bibnamefont {Zhang}},
  \bibinfo {author} {\bibfnamefont {X.}~\bibnamefont {Feng}}, \bibinfo {author}
  {\bibfnamefont {J.}~\bibnamefont {Shen}}, \bibinfo {author} {\bibfnamefont
  {Z.}~\bibnamefont {Zhang}}, \bibinfo {author} {\bibfnamefont
  {M.}~\bibnamefont {Guo}}, \bibinfo {author} {\bibfnamefont {K.}~\bibnamefont
  {Li}}, \bibinfo {author} {\bibfnamefont {Y.}~\bibnamefont {Ou}}, \bibinfo
  {author} {\bibfnamefont {P.}~\bibnamefont {Wei}}, \bibinfo {author}
  {\bibfnamefont {L.-L.}\ \bibnamefont {Wang}},  \emph {et~al.},\ }\href
  {\doibase https://doi.org/10.1126/science.1234414} {\bibfield  {journal}
  {\bibinfo  {journal} {Science}\ }\textbf {\bibinfo {volume} {340}},\ \bibinfo
  {pages} {167} (\bibinfo {year} {2013})}\BibitemShut {NoStop}%
\bibitem [{\citenamefont {Chang}\ \emph {et~al.}(2023)\citenamefont {Chang},
  \citenamefont {Liu},\ and\ \citenamefont {MacDonald}}]{chang2023colloquium}%
  \BibitemOpen
  \bibfield  {author} {\bibinfo {author} {\bibfnamefont {C.-Z.}\ \bibnamefont
  {Chang}}, \bibinfo {author} {\bibfnamefont {C.-X.}\ \bibnamefont {Liu}}, \
  and\ \bibinfo {author} {\bibfnamefont {A.~H.}\ \bibnamefont {MacDonald}},\
  }\href {\doibase https://doi.org/10.1103/RevModPhys.95.011002} {\bibfield
  {journal} {\bibinfo  {journal} {Reviews of Modern Physics}\ }\textbf
  {\bibinfo {volume} {95}},\ \bibinfo {pages} {011002} (\bibinfo {year}
  {2023})}\BibitemShut {NoStop}%
\bibitem [{\citenamefont {He}\ \emph {et~al.}(2015)\citenamefont {He},
  \citenamefont {Wang},\ and\ \citenamefont {Xue}}]{he2015quantum}%
  \BibitemOpen
  \bibfield  {author} {\bibinfo {author} {\bibfnamefont {K.}~\bibnamefont
  {He}}, \bibinfo {author} {\bibfnamefont {Y.}~\bibnamefont {Wang}}, \ and\
  \bibinfo {author} {\bibfnamefont {Q.}~\bibnamefont {Xue}},\ }\href@noop {}
  {\bibfield  {journal} {\bibinfo  {journal} {Topological Insulators:
  Fundamentals and Perspectives}\ ,\ \bibinfo {pages} {357}} (\bibinfo {year}
  {2015})}\BibitemShut {NoStop}%
\bibitem [{\citenamefont {Ou}\ \emph {et~al.}(2018)\citenamefont {Ou},
  \citenamefont {Liu}, \citenamefont {Jiang}, \citenamefont {Feng},
  \citenamefont {Zhao}, \citenamefont {Wu}, \citenamefont {Wang}, \citenamefont
  {Li}, \citenamefont {Song}, \citenamefont {Wang} \emph
  {et~al.}}]{ou2018enhancing}%
  \BibitemOpen
  \bibfield  {author} {\bibinfo {author} {\bibfnamefont {Y.}~\bibnamefont
  {Ou}}, \bibinfo {author} {\bibfnamefont {C.}~\bibnamefont {Liu}}, \bibinfo
  {author} {\bibfnamefont {G.}~\bibnamefont {Jiang}}, \bibinfo {author}
  {\bibfnamefont {Y.}~\bibnamefont {Feng}}, \bibinfo {author} {\bibfnamefont
  {D.}~\bibnamefont {Zhao}}, \bibinfo {author} {\bibfnamefont {W.}~\bibnamefont
  {Wu}}, \bibinfo {author} {\bibfnamefont {X.-X.}\ \bibnamefont {Wang}},
  \bibinfo {author} {\bibfnamefont {W.}~\bibnamefont {Li}}, \bibinfo {author}
  {\bibfnamefont {C.}~\bibnamefont {Song}}, \bibinfo {author} {\bibfnamefont
  {L.-L.}\ \bibnamefont {Wang}},  \emph {et~al.},\ }\href {\doibase
  https://doi.org/10.1002/adma.201703062} {\bibfield  {journal} {\bibinfo
  {journal} {Advanced materials}\ }\textbf {\bibinfo {volume} {30}},\ \bibinfo
  {pages} {1703062} (\bibinfo {year} {2018})}\BibitemShut {NoStop}%
\bibitem [{\citenamefont {Mogi}\ \emph {et~al.}(2015)\citenamefont {Mogi},
  \citenamefont {Yoshimi}, \citenamefont {Tsukazaki}, \citenamefont {Yasuda},
  \citenamefont {Kozuka}, \citenamefont {Takahashi}, \citenamefont {Kawasaki},\
  and\ \citenamefont {Tokura}}]{mogi2015magnetic}%
  \BibitemOpen
  \bibfield  {author} {\bibinfo {author} {\bibfnamefont {M.}~\bibnamefont
  {Mogi}}, \bibinfo {author} {\bibfnamefont {R.}~\bibnamefont {Yoshimi}},
  \bibinfo {author} {\bibfnamefont {A.}~\bibnamefont {Tsukazaki}}, \bibinfo
  {author} {\bibfnamefont {K.}~\bibnamefont {Yasuda}}, \bibinfo {author}
  {\bibfnamefont {Y.}~\bibnamefont {Kozuka}}, \bibinfo {author} {\bibfnamefont
  {K.}~\bibnamefont {Takahashi}}, \bibinfo {author} {\bibfnamefont
  {M.}~\bibnamefont {Kawasaki}}, \ and\ \bibinfo {author} {\bibfnamefont
  {Y.}~\bibnamefont {Tokura}},\ }\href {\doibase
  https://doi.org/10.1063/1.4935075} {\bibfield  {journal} {\bibinfo  {journal}
  {Applied Physics Letters}\ }\textbf {\bibinfo {volume} {107}},\ \bibinfo
  {pages} {182401} (\bibinfo {year} {2015})}\BibitemShut {NoStop}%
\bibitem [{\citenamefont {Chang}\ \emph {et~al.}(2015)\citenamefont {Chang},
  \citenamefont {Zhao}, \citenamefont {Kim}, \citenamefont {Zhang},
  \citenamefont {Assaf}, \citenamefont {Heiman}, \citenamefont {Zhang},
  \citenamefont {Liu}, \citenamefont {Chan},\ and\ \citenamefont
  {Moodera}}]{chang2015high}%
  \BibitemOpen
  \bibfield  {author} {\bibinfo {author} {\bibfnamefont {C.-Z.}\ \bibnamefont
  {Chang}}, \bibinfo {author} {\bibfnamefont {W.}~\bibnamefont {Zhao}},
  \bibinfo {author} {\bibfnamefont {D.~Y.}\ \bibnamefont {Kim}}, \bibinfo
  {author} {\bibfnamefont {H.}~\bibnamefont {Zhang}}, \bibinfo {author}
  {\bibfnamefont {B.~A.}\ \bibnamefont {Assaf}}, \bibinfo {author}
  {\bibfnamefont {D.}~\bibnamefont {Heiman}}, \bibinfo {author} {\bibfnamefont
  {S.-C.}\ \bibnamefont {Zhang}}, \bibinfo {author} {\bibfnamefont
  {C.}~\bibnamefont {Liu}}, \bibinfo {author} {\bibfnamefont {M.~H.}\
  \bibnamefont {Chan}}, \ and\ \bibinfo {author} {\bibfnamefont {J.~S.}\
  \bibnamefont {Moodera}},\ }\href {\doibase
  http://www.nature.com/doifinder/10.1038/nmat4204} {\bibfield  {journal}
  {\bibinfo  {journal} {Nature materials}\ }\textbf {\bibinfo {volume} {14}},\
  \bibinfo {pages} {473} (\bibinfo {year} {2015})}\BibitemShut {NoStop}%
\bibitem [{\citenamefont {Gong}\ \emph {et~al.}(2019)\citenamefont {Gong},
  \citenamefont {Guo}, \citenamefont {Li}, \citenamefont {Zhu}, \citenamefont
  {Liao}, \citenamefont {Liu}, \citenamefont {Zhang}, \citenamefont {Gu},
  \citenamefont {Tang}, \citenamefont {Feng} \emph
  {et~al.}}]{gong2019experimental}%
  \BibitemOpen
  \bibfield  {author} {\bibinfo {author} {\bibfnamefont {Y.}~\bibnamefont
  {Gong}}, \bibinfo {author} {\bibfnamefont {J.}~\bibnamefont {Guo}}, \bibinfo
  {author} {\bibfnamefont {J.}~\bibnamefont {Li}}, \bibinfo {author}
  {\bibfnamefont {K.}~\bibnamefont {Zhu}}, \bibinfo {author} {\bibfnamefont
  {M.}~\bibnamefont {Liao}}, \bibinfo {author} {\bibfnamefont {X.}~\bibnamefont
  {Liu}}, \bibinfo {author} {\bibfnamefont {Q.}~\bibnamefont {Zhang}}, \bibinfo
  {author} {\bibfnamefont {L.}~\bibnamefont {Gu}}, \bibinfo {author}
  {\bibfnamefont {L.}~\bibnamefont {Tang}}, \bibinfo {author} {\bibfnamefont
  {X.}~\bibnamefont {Feng}},  \emph {et~al.},\ }\href {\doibase
  https://doi.org/10.1088/0256-307X/36/7/076801} {\bibfield  {journal}
  {\bibinfo  {journal} {Chinese Physics Letters}\ }\textbf {\bibinfo {volume}
  {36}},\ \bibinfo {pages} {076801} (\bibinfo {year} {2019})}\BibitemShut
  {NoStop}%
\bibitem [{\citenamefont {Zhu}\ \emph {et~al.}(2018)\citenamefont {Zhu},
  \citenamefont {Meng}, \citenamefont {Liang}, \citenamefont {Shi},
  \citenamefont {Zhao}, \citenamefont {Cheng}, \citenamefont {Li},
  \citenamefont {Zhai}, \citenamefont {Lu}, \citenamefont {Chen} \emph
  {et~al.}}]{zhu2018proximity}%
  \BibitemOpen
  \bibfield  {author} {\bibinfo {author} {\bibfnamefont {S.}~\bibnamefont
  {Zhu}}, \bibinfo {author} {\bibfnamefont {D.}~\bibnamefont {Meng}}, \bibinfo
  {author} {\bibfnamefont {G.}~\bibnamefont {Liang}}, \bibinfo {author}
  {\bibfnamefont {G.}~\bibnamefont {Shi}}, \bibinfo {author} {\bibfnamefont
  {P.}~\bibnamefont {Zhao}}, \bibinfo {author} {\bibfnamefont {P.}~\bibnamefont
  {Cheng}}, \bibinfo {author} {\bibfnamefont {Y.}~\bibnamefont {Li}}, \bibinfo
  {author} {\bibfnamefont {X.}~\bibnamefont {Zhai}}, \bibinfo {author}
  {\bibfnamefont {Y.}~\bibnamefont {Lu}}, \bibinfo {author} {\bibfnamefont
  {L.}~\bibnamefont {Chen}},  \emph {et~al.},\ }\href {\doibase
  https://doi.org/10.1039/C8NR02083C} {\bibfield  {journal} {\bibinfo
  {journal} {Nanoscale}\ }\textbf {\bibinfo {volume} {10}},\ \bibinfo {pages}
  {10041} (\bibinfo {year} {2018})}\BibitemShut {NoStop}%
\bibitem [{\citenamefont {Adagideli}\ \emph {et~al.}(2020)\citenamefont
  {Adagideli}, \citenamefont {Hassler}, \citenamefont {Grabsch}, \citenamefont
  {Pacholski},\ and\ \citenamefont {Beenakker}}]{adagideli2020time}%
  \BibitemOpen
  \bibfield  {author} {\bibinfo {author} {\bibfnamefont {I.}~\bibnamefont
  {Adagideli}}, \bibinfo {author} {\bibfnamefont {F.}~\bibnamefont {Hassler}},
  \bibinfo {author} {\bibfnamefont {A.}~\bibnamefont {Grabsch}}, \bibinfo
  {author} {\bibfnamefont {M.}~\bibnamefont {Pacholski}}, \ and\ \bibinfo
  {author} {\bibfnamefont {C.}~\bibnamefont {Beenakker}},\ }\href {\doibase
  http://dx.doi.org/10.21468/SciPostPhys.8.1.013} {\bibfield  {journal}
  {\bibinfo  {journal} {SciPost Physics}\ }\textbf {\bibinfo {volume} {8}},\
  \bibinfo {pages} {013} (\bibinfo {year} {2020})}\BibitemShut {NoStop}%
\bibitem [{\citenamefont {Hassler}\ \emph {et~al.}(2020)\citenamefont
  {Hassler}, \citenamefont {Grabsch}, \citenamefont {Pacholski}, \citenamefont
  {Oriekhov}, \citenamefont {Ovdat}, \citenamefont {Adagideli},\ and\
  \citenamefont {Beenakker}}]{hassler2020half}%
  \BibitemOpen
  \bibfield  {author} {\bibinfo {author} {\bibfnamefont {F.}~\bibnamefont
  {Hassler}}, \bibinfo {author} {\bibfnamefont {A.}~\bibnamefont {Grabsch}},
  \bibinfo {author} {\bibfnamefont {M.}~\bibnamefont {Pacholski}}, \bibinfo
  {author} {\bibfnamefont {D.}~\bibnamefont {Oriekhov}}, \bibinfo {author}
  {\bibfnamefont {O.}~\bibnamefont {Ovdat}}, \bibinfo {author} {\bibfnamefont
  {{\.I}.}~\bibnamefont {Adagideli}}, \ and\ \bibinfo {author} {\bibfnamefont
  {C.}~\bibnamefont {Beenakker}},\ }\href {\doibase
  https://doi.org/10.1103/PhysRevB.102.045431} {\bibfield  {journal} {\bibinfo
  {journal} {Physical Review B}\ }\textbf {\bibinfo {volume} {102}},\ \bibinfo
  {pages} {045431} (\bibinfo {year} {2020})}\BibitemShut {NoStop}%
\bibitem [{\citenamefont {Liu}\ \emph {et~al.}(2018)\citenamefont {Liu},
  \citenamefont {He}, \citenamefont {Nori} \emph {et~al.}}]{liu2018majorana}%
  \BibitemOpen
  \bibfield  {author} {\bibinfo {author} {\bibfnamefont {T.}~\bibnamefont
  {Liu}}, \bibinfo {author} {\bibfnamefont {J.~J.}\ \bibnamefont {He}},
  \bibinfo {author} {\bibfnamefont {F.}~\bibnamefont {Nori}},  \emph {et~al.},\
  }\href {\doibase https://doi.org/10.1103/PhysRevB.98.245413} {\bibfield
  {journal} {\bibinfo  {journal} {Physical Review B}\ }\textbf {\bibinfo
  {volume} {98}},\ \bibinfo {pages} {245413} (\bibinfo {year}
  {2018})}\BibitemShut {NoStop}%
\bibitem [{\citenamefont {Chen}\ \emph {et~al.}(2018)\citenamefont {Chen},
  \citenamefont {Xie}, \citenamefont {Liu}, \citenamefont {Lee},\ and\
  \citenamefont {Law}}]{chen2018quasi}%
  \BibitemOpen
  \bibfield  {author} {\bibinfo {author} {\bibfnamefont {C.-Z.}\ \bibnamefont
  {Chen}}, \bibinfo {author} {\bibfnamefont {Y.-M.}\ \bibnamefont {Xie}},
  \bibinfo {author} {\bibfnamefont {J.}~\bibnamefont {Liu}}, \bibinfo {author}
  {\bibfnamefont {P.~A.}\ \bibnamefont {Lee}}, \ and\ \bibinfo {author}
  {\bibfnamefont {K.~T.}\ \bibnamefont {Law}},\ }\href {\doibase
  https://doi.org/10.1103/PhysRevB.97.104504} {\bibfield  {journal} {\bibinfo
  {journal} {Physical Review B}\ }\textbf {\bibinfo {volume} {97}},\ \bibinfo
  {pages} {104504} (\bibinfo {year} {2018})}\BibitemShut {NoStop}%
\bibitem [{\citenamefont {Zeng}\ \emph {et~al.}(2018)\citenamefont {Zeng},
  \citenamefont {Lei}, \citenamefont {Chaudhary},\ and\ \citenamefont
  {MacDonald}}]{zeng2018quantum}%
  \BibitemOpen
  \bibfield  {author} {\bibinfo {author} {\bibfnamefont {Y.}~\bibnamefont
  {Zeng}}, \bibinfo {author} {\bibfnamefont {C.}~\bibnamefont {Lei}}, \bibinfo
  {author} {\bibfnamefont {G.}~\bibnamefont {Chaudhary}}, \ and\ \bibinfo
  {author} {\bibfnamefont {A.~H.}\ \bibnamefont {MacDonald}},\ }\href {\doibase
  https://doi.org/10.1103/PhysRevB.97.081102} {\bibfield  {journal} {\bibinfo
  {journal} {Physical Review B}\ }\textbf {\bibinfo {volume} {97}},\ \bibinfo
  {pages} {081102} (\bibinfo {year} {2018})}\BibitemShut {NoStop}%
\bibitem [{\citenamefont {Beenakker}\ \emph {et~al.}(2019)\citenamefont
  {Beenakker}, \citenamefont {Baireuther}, \citenamefont {Herasymenko},
  \citenamefont {Adagideli}, \citenamefont {Wang},\ and\ \citenamefont
  {Akhmerov}}]{beenakker2019deterministic}%
  \BibitemOpen
  \bibfield  {author} {\bibinfo {author} {\bibfnamefont {C.}~\bibnamefont
  {Beenakker}}, \bibinfo {author} {\bibfnamefont {P.}~\bibnamefont
  {Baireuther}}, \bibinfo {author} {\bibfnamefont {Y.}~\bibnamefont
  {Herasymenko}}, \bibinfo {author} {\bibfnamefont {I.}~\bibnamefont
  {Adagideli}}, \bibinfo {author} {\bibfnamefont {L.}~\bibnamefont {Wang}}, \
  and\ \bibinfo {author} {\bibfnamefont {A.}~\bibnamefont {Akhmerov}},\ }\href
  {\doibase https://doi.org/10.1103/PhysRevLett.122.146803} {\bibfield
  {journal} {\bibinfo  {journal} {Physical Review Letters}\ }\textbf {\bibinfo
  {volume} {122}},\ \bibinfo {pages} {146803} (\bibinfo {year}
  {2019})}\BibitemShut {NoStop}%
\bibitem [{\citenamefont {Yu}\ \emph {et~al.}(2010)\citenamefont {Yu},
  \citenamefont {Zhang}, \citenamefont {Zhang}, \citenamefont {Zhang},
  \citenamefont {Dai},\ and\ \citenamefont {Fang}}]{yu2010quantized}%
  \BibitemOpen
  \bibfield  {author} {\bibinfo {author} {\bibfnamefont {R.}~\bibnamefont
  {Yu}}, \bibinfo {author} {\bibfnamefont {W.}~\bibnamefont {Zhang}}, \bibinfo
  {author} {\bibfnamefont {H.-J.}\ \bibnamefont {Zhang}}, \bibinfo {author}
  {\bibfnamefont {S.-C.}\ \bibnamefont {Zhang}}, \bibinfo {author}
  {\bibfnamefont {X.}~\bibnamefont {Dai}}, \ and\ \bibinfo {author}
  {\bibfnamefont {Z.}~\bibnamefont {Fang}},\ }\href {\doibase
  https://doi.org/10.1126/science.1187485} {\bibfield  {journal} {\bibinfo
  {journal} {Science}\ }\textbf {\bibinfo {volume} {329}},\ \bibinfo {pages}
  {61} (\bibinfo {year} {2010})}\BibitemShut {NoStop}%
\bibitem [{\citenamefont {Onoda}\ and\ \citenamefont
  {Nagaosa}(2003)}]{onoda2003quantized}%
  \BibitemOpen
  \bibfield  {author} {\bibinfo {author} {\bibfnamefont {M.}~\bibnamefont
  {Onoda}}\ and\ \bibinfo {author} {\bibfnamefont {N.}~\bibnamefont
  {Nagaosa}},\ }\href {\doibase https://doi.org/10.1103/PhysRevLett.90.206601}
  {\bibfield  {journal} {\bibinfo  {journal} {Physical Review Letters}\
  }\textbf {\bibinfo {volume} {90}},\ \bibinfo {pages} {206601} (\bibinfo
  {year} {2003})}\BibitemShut {NoStop}%
\bibitem [{\citenamefont {Deng}\ \emph {et~al.}(2020)\citenamefont {Deng},
  \citenamefont {Yu}, \citenamefont {Shi}, \citenamefont {Guo}, \citenamefont
  {Xu}, \citenamefont {Wang}, \citenamefont {Chen},\ and\ \citenamefont
  {Zhang}}]{deng2020quantum}%
  \BibitemOpen
  \bibfield  {author} {\bibinfo {author} {\bibfnamefont {Y.}~\bibnamefont
  {Deng}}, \bibinfo {author} {\bibfnamefont {Y.}~\bibnamefont {Yu}}, \bibinfo
  {author} {\bibfnamefont {M.~Z.}\ \bibnamefont {Shi}}, \bibinfo {author}
  {\bibfnamefont {Z.}~\bibnamefont {Guo}}, \bibinfo {author} {\bibfnamefont
  {Z.}~\bibnamefont {Xu}}, \bibinfo {author} {\bibfnamefont {J.}~\bibnamefont
  {Wang}}, \bibinfo {author} {\bibfnamefont {X.~H.}\ \bibnamefont {Chen}}, \
  and\ \bibinfo {author} {\bibfnamefont {Y.}~\bibnamefont {Zhang}},\ }\href
  {\doibase https://doi.org/10.1126/science.aax8156} {\bibfield  {journal}
  {\bibinfo  {journal} {Science}\ }\textbf {\bibinfo {volume} {367}},\ \bibinfo
  {pages} {895} (\bibinfo {year} {2020})}\BibitemShut {NoStop}%
\bibitem [{\citenamefont {Niu}\ \emph {et~al.}(2019)\citenamefont {Niu},
  \citenamefont {Mao}, \citenamefont {Hu}, \citenamefont {Huang},\ and\
  \citenamefont {Dai}}]{niu2019quantum}%
  \BibitemOpen
  \bibfield  {author} {\bibinfo {author} {\bibfnamefont {C.}~\bibnamefont
  {Niu}}, \bibinfo {author} {\bibfnamefont {N.}~\bibnamefont {Mao}}, \bibinfo
  {author} {\bibfnamefont {X.}~\bibnamefont {Hu}}, \bibinfo {author}
  {\bibfnamefont {B.}~\bibnamefont {Huang}}, \ and\ \bibinfo {author}
  {\bibfnamefont {Y.}~\bibnamefont {Dai}},\ }\href {\doibase
  https://doi.org/10.1103/PhysRevB.99.235119} {\bibfield  {journal} {\bibinfo
  {journal} {Physical Review B}\ }\textbf {\bibinfo {volume} {99}},\ \bibinfo
  {pages} {235119} (\bibinfo {year} {2019})}\BibitemShut {NoStop}%
\bibitem [{\citenamefont {Zhang}\ \emph {et~al.}(2019)\citenamefont {Zhang},
  \citenamefont {Wang}, \citenamefont {Wang}, \citenamefont {Wei},
  \citenamefont {Chen}, \citenamefont {Wang}, \citenamefont {Shi},
  \citenamefont {Wang}, \citenamefont {Jia}, \citenamefont {Ouyang} \emph
  {et~al.}}]{zhang2019experimental}%
  \BibitemOpen
  \bibfield  {author} {\bibinfo {author} {\bibfnamefont {S.}~\bibnamefont
  {Zhang}}, \bibinfo {author} {\bibfnamefont {R.}~\bibnamefont {Wang}},
  \bibinfo {author} {\bibfnamefont {X.}~\bibnamefont {Wang}}, \bibinfo {author}
  {\bibfnamefont {B.}~\bibnamefont {Wei}}, \bibinfo {author} {\bibfnamefont
  {B.}~\bibnamefont {Chen}}, \bibinfo {author} {\bibfnamefont {H.}~\bibnamefont
  {Wang}}, \bibinfo {author} {\bibfnamefont {G.}~\bibnamefont {Shi}}, \bibinfo
  {author} {\bibfnamefont {F.}~\bibnamefont {Wang}}, \bibinfo {author}
  {\bibfnamefont {B.}~\bibnamefont {Jia}}, \bibinfo {author} {\bibfnamefont
  {Y.}~\bibnamefont {Ouyang}},  \emph {et~al.},\ }\href {\doibase
  https://doi.org/10.1021/acs.nanolett.9b04555} {\bibfield  {journal} {\bibinfo
   {journal} {Nano Letters}\ }\textbf {\bibinfo {volume} {20}},\ \bibinfo
  {pages} {709} (\bibinfo {year} {2019})}\BibitemShut {NoStop}%
\bibitem [{\citenamefont {Laha}\ \emph {et~al.}(2020)\citenamefont {Laha},
  \citenamefont {Mardanya}, \citenamefont {Singh}, \citenamefont {Lin},
  \citenamefont {Bansil}, \citenamefont {Agarwal},\ and\ \citenamefont
  {Hossain}}]{laha2020magnetotransport}%
  \BibitemOpen
  \bibfield  {author} {\bibinfo {author} {\bibfnamefont {A.}~\bibnamefont
  {Laha}}, \bibinfo {author} {\bibfnamefont {S.}~\bibnamefont {Mardanya}},
  \bibinfo {author} {\bibfnamefont {B.}~\bibnamefont {Singh}}, \bibinfo
  {author} {\bibfnamefont {H.}~\bibnamefont {Lin}}, \bibinfo {author}
  {\bibfnamefont {A.}~\bibnamefont {Bansil}}, \bibinfo {author} {\bibfnamefont
  {A.}~\bibnamefont {Agarwal}}, \ and\ \bibinfo {author} {\bibfnamefont
  {Z.}~\bibnamefont {Hossain}},\ }\href {\doibase
  https://doi.org/10.1103/PhysRevB.102.035164} {\bibfield  {journal} {\bibinfo
  {journal} {Physical Review B}\ }\textbf {\bibinfo {volume} {102}},\ \bibinfo
  {pages} {035164} (\bibinfo {year} {2020})}\BibitemShut {NoStop}%
\bibitem [{\citenamefont {Cao}\ \emph {et~al.}(2021)\citenamefont {Cao},
  \citenamefont {Han}, \citenamefont {Lv}, \citenamefont {Wang}, \citenamefont
  {Luo}, \citenamefont {Zhang}, \citenamefont {Yao}, \citenamefont {Zhou},
  \citenamefont {Chen}, \citenamefont {Zhang} \emph {et~al.}}]{cao2021growth}%
  \BibitemOpen
  \bibfield  {author} {\bibinfo {author} {\bibfnamefont {L.}~\bibnamefont
  {Cao}}, \bibinfo {author} {\bibfnamefont {S.}~\bibnamefont {Han}}, \bibinfo
  {author} {\bibfnamefont {Y.-Y.}\ \bibnamefont {Lv}}, \bibinfo {author}
  {\bibfnamefont {D.}~\bibnamefont {Wang}}, \bibinfo {author} {\bibfnamefont
  {Y.-C.}\ \bibnamefont {Luo}}, \bibinfo {author} {\bibfnamefont {Y.-Y.}\
  \bibnamefont {Zhang}}, \bibinfo {author} {\bibfnamefont {S.-H.}\ \bibnamefont
  {Yao}}, \bibinfo {author} {\bibfnamefont {J.}~\bibnamefont {Zhou}}, \bibinfo
  {author} {\bibfnamefont {Y.}~\bibnamefont {Chen}}, \bibinfo {author}
  {\bibfnamefont {H.}~\bibnamefont {Zhang}},  \emph {et~al.},\ }\href {\doibase
  https://doi.org/10.1103/PhysRevB.104.054421} {\bibfield  {journal} {\bibinfo
  {journal} {Physical Review B}\ }\textbf {\bibinfo {volume} {104}},\ \bibinfo
  {pages} {054421} (\bibinfo {year} {2021})}\BibitemShut {NoStop}%
\bibitem [{\citenamefont {Watson}\ \emph {et~al.}(2015)\citenamefont {Watson},
  \citenamefont {Yamashita}, \citenamefont {Kasahara}, \citenamefont {Knafo},
  \citenamefont {Nardone}, \citenamefont {B{\'e}ard}, \citenamefont {Hardy},
  \citenamefont {McCollam}, \citenamefont {Narayanan}, \citenamefont {Blake}
  \emph {et~al.}}]{watson2015dichotomy}%
  \BibitemOpen
  \bibfield  {author} {\bibinfo {author} {\bibfnamefont {M.}~\bibnamefont
  {Watson}}, \bibinfo {author} {\bibfnamefont {T.}~\bibnamefont {Yamashita}},
  \bibinfo {author} {\bibfnamefont {S.}~\bibnamefont {Kasahara}}, \bibinfo
  {author} {\bibfnamefont {W.}~\bibnamefont {Knafo}}, \bibinfo {author}
  {\bibfnamefont {M.}~\bibnamefont {Nardone}}, \bibinfo {author} {\bibfnamefont
  {J.}~\bibnamefont {B{\'e}ard}}, \bibinfo {author} {\bibfnamefont
  {F.}~\bibnamefont {Hardy}}, \bibinfo {author} {\bibfnamefont
  {A.}~\bibnamefont {McCollam}}, \bibinfo {author} {\bibfnamefont
  {A.}~\bibnamefont {Narayanan}}, \bibinfo {author} {\bibfnamefont
  {S.}~\bibnamefont {Blake}},  \emph {et~al.},\ }\href {\doibase
  http://dx.doi.org/10.1103/PhysRevLett.115.027006} {\bibfield  {journal}
  {\bibinfo  {journal} {Physical Review Letters}\ }\textbf {\bibinfo {volume}
  {115}},\ \bibinfo {pages} {027006} (\bibinfo {year} {2015})}\BibitemShut
  {NoStop}%
\bibitem [{\citenamefont {Budhani}\ \emph {et~al.}(2021)\citenamefont
  {Budhani}, \citenamefont {Higgins}, \citenamefont {McAlmont},\ and\
  \citenamefont {Paglione}}]{budhani2021planar}%
  \BibitemOpen
  \bibfield  {author} {\bibinfo {author} {\bibfnamefont {R.~C.}\ \bibnamefont
  {Budhani}}, \bibinfo {author} {\bibfnamefont {J.~S.}\ \bibnamefont
  {Higgins}}, \bibinfo {author} {\bibfnamefont {D.}~\bibnamefont {McAlmont}}, \
  and\ \bibinfo {author} {\bibfnamefont {J.}~\bibnamefont {Paglione}},\ }\href
  {\doibase https://doi.org/10.1063/5.0049577} {\bibfield  {journal} {\bibinfo
  {journal} {AIP Advances}\ }\textbf {\bibinfo {volume} {11}},\ \bibinfo
  {pages} {055020} (\bibinfo {year} {2021})}\BibitemShut {NoStop}%
\bibitem [{\citenamefont {Shi}\ \emph {et~al.}(2020)\citenamefont {Shi},
  \citenamefont {Zhang}, \citenamefont {Yan}, \citenamefont {Feng},
  \citenamefont {Yang}, \citenamefont {Shi},\ and\ \citenamefont
  {Li}}]{shi2020anomalous}%
  \BibitemOpen
  \bibfield  {author} {\bibinfo {author} {\bibfnamefont {G.}~\bibnamefont
  {Shi}}, \bibinfo {author} {\bibfnamefont {M.}~\bibnamefont {Zhang}}, \bibinfo
  {author} {\bibfnamefont {D.}~\bibnamefont {Yan}}, \bibinfo {author}
  {\bibfnamefont {H.}~\bibnamefont {Feng}}, \bibinfo {author} {\bibfnamefont
  {M.}~\bibnamefont {Yang}}, \bibinfo {author} {\bibfnamefont {Y.}~\bibnamefont
  {Shi}}, \ and\ \bibinfo {author} {\bibfnamefont {Y.}~\bibnamefont {Li}},\
  }\href {\doibase https://doi.org/10.1088/0256-307X/37/4/047301} {\bibfield
  {journal} {\bibinfo  {journal} {Chinese Physics Letters}\ }\textbf {\bibinfo
  {volume} {37}},\ \bibinfo {pages} {047301} (\bibinfo {year}
  {2020})}\BibitemShut {NoStop}%
\bibitem [{\citenamefont {Ito}\ \emph {et~al.}(2022)\citenamefont {Ito},
  \citenamefont {Masutomi},\ and\ \citenamefont
  {Okamoto}}]{ito2022cancellation}%
  \BibitemOpen
  \bibfield  {author} {\bibinfo {author} {\bibfnamefont {N.}~\bibnamefont
  {Ito}}, \bibinfo {author} {\bibfnamefont {R.}~\bibnamefont {Masutomi}}, \
  and\ \bibinfo {author} {\bibfnamefont {T.}~\bibnamefont {Okamoto}},\ }\href
  {\doibase https://doi.org/10.1103/PhysRevB.105.205434} {\bibfield  {journal}
  {\bibinfo  {journal} {Physical Review B}\ }\textbf {\bibinfo {volume}
  {105}},\ \bibinfo {pages} {205434} (\bibinfo {year} {2022})}\BibitemShut
  {NoStop}%
\bibitem [{\citenamefont {Knispel}\ \emph {et~al.}(2017)\citenamefont
  {Knispel}, \citenamefont {Jolie}, \citenamefont {Borgwardt}, \citenamefont
  {Lux}, \citenamefont {Wang}, \citenamefont {Ando}, \citenamefont {Rosch},
  \citenamefont {Michely},\ and\ \citenamefont
  {Gr{\"u}ninger}}]{knispel2017charge}%
  \BibitemOpen
  \bibfield  {author} {\bibinfo {author} {\bibfnamefont {T.}~\bibnamefont
  {Knispel}}, \bibinfo {author} {\bibfnamefont {W.}~\bibnamefont {Jolie}},
  \bibinfo {author} {\bibfnamefont {N.}~\bibnamefont {Borgwardt}}, \bibinfo
  {author} {\bibfnamefont {J.}~\bibnamefont {Lux}}, \bibinfo {author}
  {\bibfnamefont {Z.}~\bibnamefont {Wang}}, \bibinfo {author} {\bibfnamefont
  {Y.}~\bibnamefont {Ando}}, \bibinfo {author} {\bibfnamefont {A.}~\bibnamefont
  {Rosch}}, \bibinfo {author} {\bibfnamefont {T.}~\bibnamefont {Michely}}, \
  and\ \bibinfo {author} {\bibfnamefont {M.}~\bibnamefont {Gr{\"u}ninger}},\
  }\href {\doibase https://doi.org/10.1103/PhysRevB.96.195135} {\bibfield
  {journal} {\bibinfo  {journal} {Physical Review B}\ }\textbf {\bibinfo
  {volume} {96}},\ \bibinfo {pages} {195135} (\bibinfo {year}
  {2017})}\BibitemShut {NoStop}%
\bibitem [{\citenamefont {Borgwardt}\ \emph {et~al.}(2016)\citenamefont
  {Borgwardt}, \citenamefont {Lux}, \citenamefont {Vergara}, \citenamefont
  {Wang}, \citenamefont {Taskin}, \citenamefont {Segawa}, \citenamefont
  {Van~Loosdrecht}, \citenamefont {Ando}, \citenamefont {Rosch},\ and\
  \citenamefont {Gr{\"u}ninger}}]{borgwardt2016self}%
  \BibitemOpen
  \bibfield  {author} {\bibinfo {author} {\bibfnamefont {N.}~\bibnamefont
  {Borgwardt}}, \bibinfo {author} {\bibfnamefont {J.}~\bibnamefont {Lux}},
  \bibinfo {author} {\bibfnamefont {I.}~\bibnamefont {Vergara}}, \bibinfo
  {author} {\bibfnamefont {Z.}~\bibnamefont {Wang}}, \bibinfo {author}
  {\bibfnamefont {A.}~\bibnamefont {Taskin}}, \bibinfo {author} {\bibfnamefont
  {K.}~\bibnamefont {Segawa}}, \bibinfo {author} {\bibfnamefont
  {P.}~\bibnamefont {Van~Loosdrecht}}, \bibinfo {author} {\bibfnamefont
  {Y.}~\bibnamefont {Ando}}, \bibinfo {author} {\bibfnamefont {A.}~\bibnamefont
  {Rosch}}, \ and\ \bibinfo {author} {\bibfnamefont {M.}~\bibnamefont
  {Gr{\"u}ninger}},\ }\href {\doibase
  http://dx.doi.org/10.1103/PhysRevB.93.245149} {\bibfield  {journal} {\bibinfo
   {journal} {Physical Review B}\ }\textbf {\bibinfo {volume} {93}},\ \bibinfo
  {pages} {245149} (\bibinfo {year} {2016})}\BibitemShut {NoStop}%
\bibitem [{\citenamefont {Zimmermann}\ \emph {et~al.}(2023)\citenamefont
  {Zimmermann}, \citenamefont {K{\"o}lzer}, \citenamefont {Schleenvoigt},
  \citenamefont {Rosenbach}, \citenamefont {Mussler}, \citenamefont
  {Sch{\"u}ffelgen}, \citenamefont {Heider}, \citenamefont {Plucinski},
  \citenamefont {Schubert}, \citenamefont {L\"uth}, \citenamefont
  {Gr\"utzmacher},\ and\ \citenamefont {Sch\"apers}}]{zimmermann2023universal}%
  \BibitemOpen
  \bibfield  {author} {\bibinfo {author} {\bibfnamefont {E.}~\bibnamefont
  {Zimmermann}}, \bibinfo {author} {\bibfnamefont {J.}~\bibnamefont
  {K{\"o}lzer}}, \bibinfo {author} {\bibfnamefont {M.}~\bibnamefont
  {Schleenvoigt}}, \bibinfo {author} {\bibfnamefont {D.}~\bibnamefont
  {Rosenbach}}, \bibinfo {author} {\bibfnamefont {G.}~\bibnamefont {Mussler}},
  \bibinfo {author} {\bibfnamefont {P.}~\bibnamefont {Sch{\"u}ffelgen}},
  \bibinfo {author} {\bibfnamefont {T.}~\bibnamefont {Heider}}, \bibinfo
  {author} {\bibfnamefont {L.}~\bibnamefont {Plucinski}}, \bibinfo {author}
  {\bibfnamefont {J.}~\bibnamefont {Schubert}}, \bibinfo {author}
  {\bibfnamefont {H.}~\bibnamefont {L\"uth}}, \bibinfo {author} {\bibfnamefont
  {D.}~\bibnamefont {Gr\"utzmacher}}, \ and\ \bibinfo {author} {\bibfnamefont
  {T.}~\bibnamefont {Sch\"apers}},\ }\href {\doibase
  https://doi.org/10.1088/1361-6641/acb45f} {\bibfield  {journal} {\bibinfo
  {journal} {Semiconductor Science and Technology}\ }\textbf {\bibinfo {volume}
  {38}},\ \bibinfo {pages} {035010} (\bibinfo {year} {2023})}\BibitemShut
  {NoStop}%
\bibitem [{Note1()}]{Note1}%
  \BibitemOpen
  \bibinfo {note} {The data shown here is actual data from an MTI Hall
  bar.}\BibitemShut {Stop}%
\bibitem [{\citenamefont {Ye}\ \emph {et~al.}(1999)\citenamefont {Ye},
  \citenamefont {Kim}, \citenamefont {Millis}, \citenamefont {Shraiman},
  \citenamefont {Majumdar},\ and\ \citenamefont
  {Te{\v{s}}anovi{\'c}}}]{ye1999berry}%
  \BibitemOpen
  \bibfield  {author} {\bibinfo {author} {\bibfnamefont {J.}~\bibnamefont
  {Ye}}, \bibinfo {author} {\bibfnamefont {Y.~B.}\ \bibnamefont {Kim}},
  \bibinfo {author} {\bibfnamefont {A.}~\bibnamefont {Millis}}, \bibinfo
  {author} {\bibfnamefont {B.}~\bibnamefont {Shraiman}}, \bibinfo {author}
  {\bibfnamefont {P.}~\bibnamefont {Majumdar}}, \ and\ \bibinfo {author}
  {\bibfnamefont {Z.}~\bibnamefont {Te{\v{s}}anovi{\'c}}},\ }\href {\doibase
  https://doi.org/10.1103/PhysRevLett.83.3737} {\bibfield  {journal} {\bibinfo
  {journal} {Physical Review Letters}\ }\textbf {\bibinfo {volume} {83}},\
  \bibinfo {pages} {3737} (\bibinfo {year} {1999})}\BibitemShut {NoStop}%
\bibitem [{\citenamefont {Furdyna}(1988)}]{furdyna1988diluted}%
  \BibitemOpen
  \bibfield  {author} {\bibinfo {author} {\bibfnamefont {J.~K.}\ \bibnamefont
  {Furdyna}},\ }\href {\doibase https://doi.org/10.1063/1.341700} {\bibfield
  {journal} {\bibinfo  {journal} {Journal of Applied Physics}\ }\textbf
  {\bibinfo {volume} {64}},\ \bibinfo {pages} {R29} (\bibinfo {year}
  {1988})}\BibitemShut {NoStop}%
\bibitem [{\citenamefont {Albrecht}\ \emph {et~al.}(2017)\citenamefont
  {Albrecht}, \citenamefont {Moers},\ and\ \citenamefont
  {Hermanns}}]{albrecht2017hnf}%
  \BibitemOpen
  \bibfield  {author} {\bibinfo {author} {\bibfnamefont {W.}~\bibnamefont
  {Albrecht}}, \bibinfo {author} {\bibfnamefont {J.}~\bibnamefont {Moers}}, \
  and\ \bibinfo {author} {\bibfnamefont {B.}~\bibnamefont {Hermanns}},\ }\href
  {http://dx.doi.org/10.17815/jlsrf-3-158} {\bibfield  {journal} {\bibinfo
  {journal} {Journal of large-scale research facilities JLSRF}\ }\textbf
  {\bibinfo {volume} {3}},\ \bibinfo {pages} {112} (\bibinfo {year}
  {2017})}\BibitemShut {NoStop}%
\end{thebibliography}

%

\clearpage
\widetext

\setcounter{section}{0}
\setcounter{equation}{0}
\setcounter{figure}{0}
\setcounter{table}{0}
\setcounter{page}{1}
\makeatletter
\renewcommand{\thesection}{S\Roman{section}}
\renewcommand{\thesubsection}{\Alph{subsection}}
\renewcommand{\theequation}{S\arabic{equation}}
\renewcommand{\thefigure}{S\arabic{figure}}
\renewcommand{\figurename}{Supplementary Figure}
\renewcommand{\bibnumfmt}[1]{[S#1]}
\renewcommand{\citenumfont}[1]{S#1}

\begin{center}
\textbf{Supplementary Material: Fourier transformation based analysis routine for intermixed longitudinal and transversal hysteretic data for the example of a magnetic topological insulator}
\end{center}

\section{Fabrication and measurement technique}
The magnetic topological insulator (MTI) material is grown on a Si(111) substrate. First, the substrate is cleaned with a Piranha solution followed by 1\% hydrofluoric acid, which removes the native SiO$_2$ and passivates the Si(111) surface. Then, the chip is loaded into a molecular beam epitaxy (MBE) chamber and heated up to $700^\circ$C in order to desorb the hydrogen passivation. Utilizing a flux ratio of 2:9:80 (Bi:Sb:Te), a $8.6\,$nm thick ternary TI thin film is grown at a substrate temperature of $230\,^\circ$C, which is continuously doped by simultaneous Cr evaporation during growth. Rutherford backscattering spectroscopy yields a stoichiometry of Cr$_{0.15}$Bi$_{0.35}$Sb$_{1.5}$Te$_{3}$, as shown in the next section. The chromium doped MTI grows in quintuple layers where the chromium locates at bismuth/antimony atom positions. In the end, the sample is capped under in-situ conditions with $5\,$nm of Al$_2$O$_3$ as a capping layer that prevents the sample from oxidation. Using standard optical lithography processes the MTI is etched into a Hall bar shape using reactive ion etching (RIE) and contacted with $50\,$nm Ti/$100\,$nm Pt metal contacts via sputtering. A dielectric layer of $15\,$nm HfO$_2$ is deposited by atomic layer deposition (ALD) and a $50\,$nm Ti/$100\,$nm Au gate is placed on top.\\
Figure~2 and~3 in the main text are used to explain the symmetries of MTIs. Actually, real data from a similar reference sample that does not show any intermixing is used in order to produce these plots.\\
The asymmetric magnetotransport data is obtained from a $6\,\mu$m wide and $300\,\mu$m long MTI Hall bar. The applied current used for all measurements is $I=1\,\mu$A. The measurements are conducted in a variable temperature insert (VTI) cryostat with a base temperature of $1.3\,$K and a superconducting magnet is used to apply a magnetic field up to $14\,$T. All measurements are performed in a four-terminal setup so that contact resistances are not measured. 

\section{Rutherford backscattering spectrometry}
\begin{figure}[hb]
\centering
\includegraphics[width=0.7\textwidth]{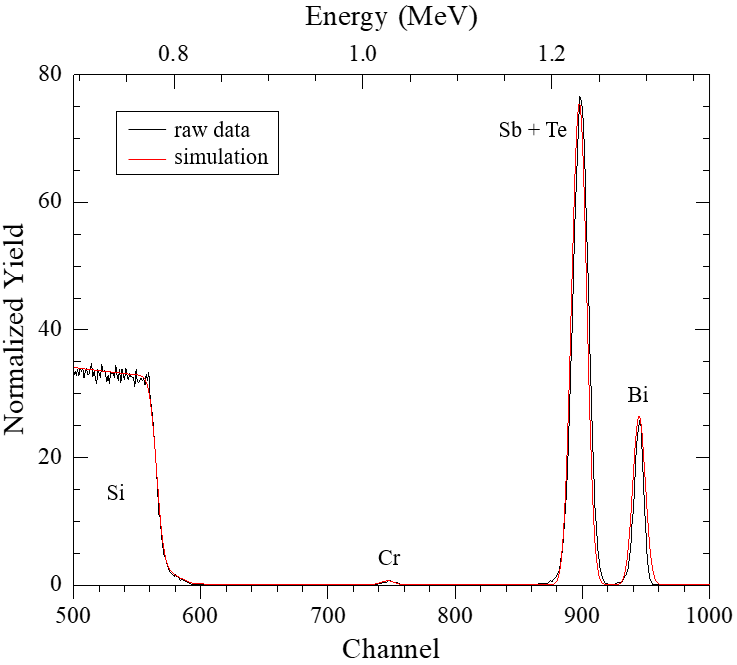}
\caption{RBS spectrum of the sample. The peaks are labeled with the corresponding material. In red the simulated curve of the determined stoichiometry is given.}\label{Fig_Supp_RBS}
\end{figure}

The analysis of the stoichiometry is done via Rutherford backscattering spectrometry (RBS) measurements. The data is shown in Fig.~\ref{Fig_Supp_RBS}. Due to their similar atomic number the peaks of antimony and tellurium lie close to each other resulting in a single peak. Knowing that the ratio of tellurium compared to the other elements is 3:2 due to the regular growth in quintuple layers the missing numbers can be calculated. The stoichiometry is determined to Cr$_{0.15}$Bi$_{0.35}$Sb$_{1.5}$Te$_{3}$ and verified with a simulation that is plotted in addition to the data in Fig.~\ref{Fig_Supp_RBS}.

\section{Symmetrization method}
Axial symmetry is mathematically described with the relation $R(B)=R(-B)$ while point symmetric functions follow $R(B)=-R(-B)$. Using the concept that every non-symmetric analytic function can be split into series of axial and point symmetric subfunctions one can derive the unperturbed longitudinal and Hall signal by symmetrizing and anti-symmetrizing the intermixed data: $R_\text{xx}=(R_\text{xx}(B)+R_\text{xx}(-B))/2$; $R_\text{xy}=(R_\text{xy}(B)-R_\text{xy}(-B))/2$. As equivalent option, in the main text a fast Fourier transformation is selected, where the odd or even contributions are omitted, respectively. This method has the additional advantage that the contributions are further divided with respect to their frequency and thus an insight of the spectrum is gained. This may be helpful for instance when symmetrizing periodic data.

\section{Filtered signal}
\begin{figure}[hbtp]
\centering
\includegraphics[width=0.7\textwidth]{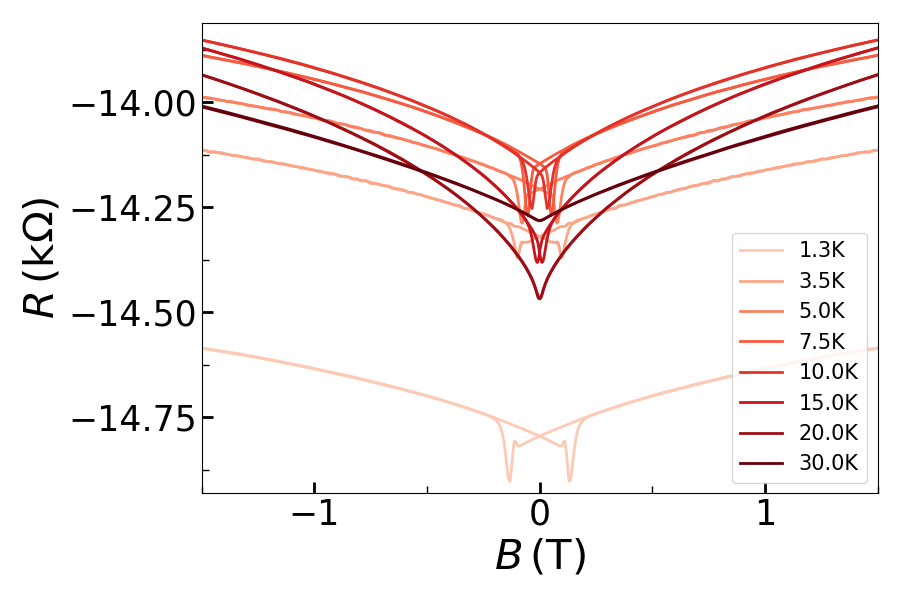}
\caption{Excluded axial symmetric part from the raw data shown in Fig.~5~b). The remaining point symmetric signal is shown in Fig.~6~b).}\label{Fig_Supp_Rxxxy}
\end{figure}

In the main article, Fig.~6~b) shows the point symmetric part of the signal after omitting the axial symmetric part using FFT. In Fig.~\ref{Fig_Supp_Rxxxy} the axial symmetric part that is excluded is shown. Here, one can clearly see that it is based on the influence of the (negative) longitudinal signal.

\section{Mathematical description of the hysteresis}
For the mathematical description of the hysteresis in the main article a model based on the error function is introduced in equation~1. One can see in Fig.~6~c) that the model describes the hysteresis in our case quite well for low temperatures but the curvature is not described exactly by the error function for elevated temperatures. There are two main reasons that cause the data to differ from the optimal Heaviside step function: First, the magnetic moments introduced by the chromium doping have slightly different switching energies and thus also slightly different coercive magnetic fields because moments close to the surface are less influenced by other surrounding magnetic moments. Furthermore, the switching energy also depends on the homogeneity of the sample, especially the homogeneity of the distribution of magnetic atoms. This distribution in energy is approximately described by a Gaussian function and thus the error function is a reasonable choice.\\
A second parameter that can cause spontaneous switching of the orientation of a magnetic moment is temperature. This effect is best described by Fermi-Dirac statistics. A description of the data using the model presented in the main article is sufficient for the determination of the anomalous hysteresis height $R_\text{AH}$ and coercive magnetic field $B_c$ even at elevated temperatures. If exact information about the curvature at elevated temperatures is needed, a combination of both, error function and Fermi–Dirac distribution, is suggested. As the article addresses not only a magnetic topological insulator-based hysteresis other physical concepts may be additionally needed to be taken into account when searching for a sufficient mathematical description.

\end{document}